# Site-controlled formation of single Si nanocrystals in a buried SiO$_2$ matrix using ion beam mixing

Xiaomo Xu[1], Thomas Prüfer[1], Daniel Wolf[1,2], Hans-Jürgen Engelmann[1], Lothar Bischoff[1], René Hübner[1], Karl-Heinz Heinig[1], Wolfhard Möller[1], Stefan Facsko[1], Johannes von Borany[1] and Gregor Hlawacek[*1]

## Abstract

For future nanoelectronic devices – such as room-temperature single electron transistors – the site-controlled formation of single Si nanocrystals (NCs) is a crucial prerequisite. Here, we report an approach to fabricate single Si NCs via medium-energy Si$^+$ or Ne$^+$ ion beam mixing of Si into a buried SiO$_2$ layer followed by thermally activated phase separation. Binary collision approximation and kinetic Monte Carlo methods are conducted to gain atomistic insight into the influence of relevant experimental parameters on the Si NC formation process. Energy-filtered transmission electron microscopy is performed to obtain quantitative values on the Si NC size and distribution in dependence of the layer stack geometry, ion fluence and thermal budget. Employing a focused Ne$^+$ beam from a helium ion microscope, we demonstrate site-controlled self-assembly of single Si NCs. Line irradiation with a fluence of 3000 Ne$^+$/nm$^2$ and a line width of 4 nm leads to the formation of a chain of Si NCs, and a single NC with 2.2 nm diameter is subsequently isolated and visualized in a few nanometer thin lamella prepared by a focused ion beam (FIB). The Si NC is centered between the SiO$_2$ layers and perpendicular to the incident Ne$^+$ beam.

## Introduction
Silicon has been the main material in the semiconductor industry for almost all use cases with the exception of optical applications. The latter is due to its indirect band gap in the bulk state. Benefiting from their reduced size, Si NCs show optical activity [1,2] and quantum confinement behavior [3] and have inspired novel applications in microelectronics [4], optics [1] and photovoltaics [5,6]. In particular, various groups have demonstrated the usage of a Si NC embedded in an SiO$_2$ matrix as a Coulomb island for a single electron transistor (SET) device [7-9]. However, so far Si NC-based SET devices lack either the ability of room-temperature operation or the compatibility to complimentary metal-oxide-semiconductor (CMOS) technology and thus have yet to be integrated into a cost-efficient Si-based technology.

Multiple methods have been proposed and optimized for Si NC fabrication, including plasma-enhanced chemical vapor deposition (PECVD) [4,10], magnetron sputtering [11,12], laser-induced pyrolysis [13] and ion beam synthesis in an $SiO_2$ matrix [14-17]. Compared to conventional ion beam synthesis using low-energy ion implantation, collisional mixing of Si into an $SiO_2$ layer by ion irradiation at higher energies leads to a better control over the Si excess, and a self-aligned δ-layer of NCs can form near the Si/$SiO_2$ interface after a post-annealing process [18]. Si NC layer formation via ion beam mixing, which is studied both theoretically and experimentally [19], has capacitated the manufacturing of Si NC-based non-volatile memory devices [20].

A possible route to fabricate a Si NC-based SET device is to obtain lateral control over the formation of few or even a single Si NC using the ion beam approach. To obtain an estimate for the charging energy of the Coulomb island $E_c = \frac{e^2}{C}$, we use the self capacitance of a sphere $C = 4\pi\varepsilon\varepsilon_0 r_{NC}$. In order to have $E_c$ larger than 5 kT at room temperature, the diameter of an individual Si NC, $2r_{NC}$, has to be smaller than 5.7 nm. The other factors are the unit charge $e$ and $\varepsilon_0$ and $\varepsilon_r = 3.9$ are the vacuum permittivity and the relative permittivity of silicon dioxide, respectively. Recently, advanced lithographic methods [21] and directed self-assembly (DSA) techniques [22] have been implemented to achieve a small implanted volume and consequently a small number of Si NCs. However, neither approach broke through the limit to achieve a volume small enough for single Si NC formation. To achieve this goal, we demonstrate the usage of a FIB for laterally precise irradiation. The recent advance in noble gas ion microscopy, in particular the availability of a highly focused $Ne^+$ beam from a helium ion microscope (HIM), provides ultimate control over the irradiation geometry and fluence [23,24], which leads to a minimal mixed volume and the formation of a single Si NC. This proof-of-principle approach opens the possibility for future investigation of Si NC-based SET fabrication.

In this paper, first the formation of a Si NC δ-layer in Si/$SiO_2$/Si stacks by broad-beam $Si^+$ irradiation is studied for different buried oxide thicknesses. This experimental work is supported by computer simulations of the ion beam mixing and phase separation process. Second, a systematic study of $Si^+$ NC formation is reported, to define optimized irradiation and annealing parameters. Third, these parameters are adapted to the FIB approach using the HIM. For this scenario, dynamic binary collision and kinetic Monte Carlo (kMC) computer simulations predict the formation of a chain of single Si NCs due to the strong reduction of the $SiO_2$ volume in which the Si excess is sufficient for NC formation. Finally, a single Si NC embedded in $SiO_2$ is isolated by FIB-based transmission electron microscopy (TEM) lamella preparation. In all cases, the local distribution and size of the NCs are mapped using energy-filtered transmission electron microscopy (EFTEM).

## Results and Discussion

In Figure 1, a comparison of cross-sectional Si plasmon-loss-filtered TEM images obtained from two Si/$SiO_2$/Si layer stacks with different oxide layer thickness of 14 nm (Figure 1a) and 7 nm (Figure 1b) is presented. The samples have been irradiated with $Si^+$ ions using a broad beam.

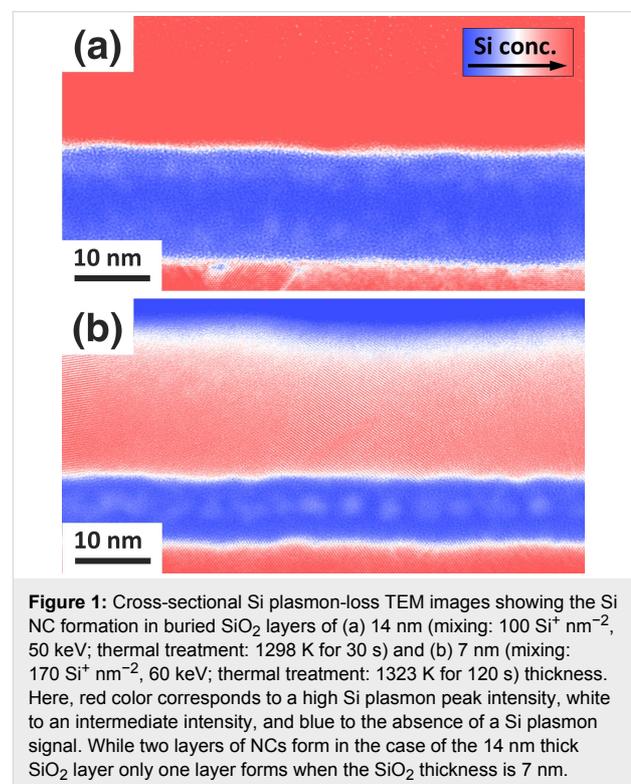

**Figure 1:** Cross-sectional Si plasmon-loss TEM images showing the Si NC formation in buried $SiO_2$ layers of (a) 14 nm (mixing: 100 $Si^+$ $nm^{-2}$, 50 keV; thermal treatment: 1298 K for 30 s) and (b) 7 nm (mixing: 170 $Si^+$ $nm^{-2}$, 60 keV; thermal treatment: 1323 K for 120 s) thickness. Here, red color corresponds to a high Si plasmon peak intensity, white to an intermediate intensity, and blue to the absence of a Si plasmon signal. While two layers of NCs form in the case of the 14 nm thick $SiO_2$ layer only one layer forms when the $SiO_2$ thickness is 7 nm.

The irradiation conditions differ to compensate for the different sample geometries, but have been chosen to achieve comparable mixing profiles. For the 14 nm thick oxide layer, a fluence of 100 $Si^+$ $nm^{-2}$ with an energy of 50 keV and thermal treatment conditions of 1298 K for 30 s has been used, whereas the sample with the 7 nm thick oxide layer was irradiated with 60 keV $Si^+$ ions and a fluence of 170 $nm^{-2}$. The thermal treatment was performed at 1323 K for 120 s.

In both cases, Si NCs (visible as white round shapes in the blue oxide layer) are formed in the buried $SiO_2$ layer. However, in case of the thinner (7 nm) oxide, the NCs are arranged in a single layer, whereas in the 14 nm thick $SiO_2$ layer, two planes of Si NCs are formed close to the Si/SiO$_2$ interfaces. The NCs in the 7 nm case have a diameter of 2.5 ± 0.4 nm. Considering the thickness of the SiO$_2$ layer, the distance between the NCs and the Si/SiO$_2$ interfaces is approximately 2 nm. Taking into account the thickness of the classically prepared TEM lamella – measured using the inelastic mean free path of the electron [25] – we estimate an average distance between the NCs of approximately 12 nm. Please note that the EFTEM image shows a projection of the NCs. Therefore, the estimated NC separation in the 2D layer is larger than the visually observed distance in the cross sectional EFTEM image.

A combination of the binary collision approximation (BCA) method using TRIDYN [26] and kMC simulations [27] was used to gain atomistic insight into the observed NC formation and its dependence on the layer thicknesses, the irradiation conditions and the thermal treatment. In Figure 2, the results are visualized for 50 keV and 60 keV Si$^+$ irradiations of thicker (14.5 nm SiO$_2$ with 50 nm top Si) and thinner (7 nm SiO$_2$ with 30 nm top Si) layer stacks, respectively.

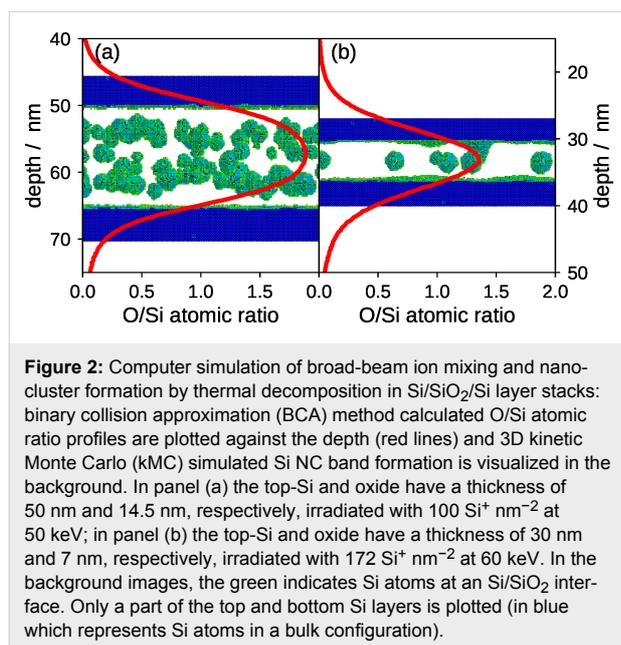

**Figure 2:** Computer simulation of broad-beam ion mixing and nanocluster formation by thermal decomposition in Si/SiO$_2$/Si layer stacks: binary collision approximation (BCA) method calculated O/Si atomic ratio profiles are plotted against the depth (red lines) and 3D kinetic Monte Carlo (kMC) simulated Si NC band formation is visualized in the background. In panel (a) the top-Si and oxide have a thickness of 50 nm and 14.5 nm, respectively, irradiated with 100 Si$^+$ nm$^{-2}$ at 50 keV; in panel (b) the top-Si and oxide have a thickness of 30 nm and 7 nm, respectively, irradiated with 172 Si$^+$ nm$^{-2}$ at 60 keV. In the background images, the green indicates Si atoms at an Si/SiO$_2$ interface. Only a part of the top and bottom Si layers is plotted (in blue which represents Si atoms in a bulk configuration).

TRIDYN simulations reveal that ion irradiation leads to interface blurring due to ion-induced atomic mixing [28], which is the result of the collisional relocation of Si atoms into the buried SiO$_2$ layer as well as the transport of oxygen atoms from the oxide into the top and bottom Si layers. The overlaying plots show the O/Si atomic ratio as a function of depth after the ion beam mixing and prior to the simulated annealing. The simulations show that after ion beam mixing with 100 Si$^+$ nm$^{-2}$ at 50 keV only a small change in stoichiometry is expected for the center of the thick oxide layer. A high Si excess due to ion beam mixing is only observed close to the Si/SiO$_2$ interface. As a result, two layers of silicon NCs form next to the interface during the phase separation. The O/Si atomic ratio is slightly larger than 1.5 at the center of the Si NC bands. The simulated distribution of Si NCs is in agreement with the experimentally observed double band of crystals in the case of the thick oxide layer (see Figure 1a). However, experimentally a more homogeneous size distribution and a clear separation in two bands of NCs is observed. In contrast, for the case of a thinner oxide, the stoichiometry does not only change close to the interfaces, but also in the middle of the oxide layer. After a fluence of 172 Si$^+$ nm$^{-2}$ at 60 keV, the O/Si atomic ratio in the center of the oxide has decreased below 1.5. After simulated annealing, a Si NC band forms close to the interface. Due to the geometric confinement only one layer of silicon NCs forms in the center of the buried oxide layer. Here, a lower density of NCs and even NCs that are connected to the interface are observed in the simulation (see Figure 2b).

In general, by carefully comparing Figure 1 with the simulated Si NC bands, presented in Figure 2, one can see that neither the cluster size nor the density matches perfectly. To obtain a comparable density and NC size, the fluence in the simulation needs to be significantly lower for all investigated cases. This systematic deviation and the underlying physics is subject of an ongoing investigation.

In order to optimize the self-assembly of vertically self-aligned Si NCs with a narrow size distribution, appropriate ion irradiation and thermal treatment conditions have to be identified. Selected examples of all investigated samples are presented in Figure 3.

Figure 3a and Figure 3b compare different ion beam mixing conditions at an identical thermal budget ($T$ = 1323 K, $t$ = 60 s). If the applied fluence is too small (see Figure 3a), insufficient mixing occurs and consequently only poorly defined NCs are formed. A too high fluence, on the other hand, will result in Si NC formation, but the interface roughness increases (see Figure 3b).

The phase separation process during post-irradiation thermal treatment is activated by the energy available for diffusion of the various constituents. The effectivity of this thermally activated process will not only depend on the temperature during the annealing process but also on the time given to the diffusing atomic species. For presentation purposes, we combine the two parameters, time and temperature, into one quantity – the thermal budget (TB) in units of cm$^2$, based on a formalism used to describe the diffusion of dopants in semiconductors [29]. We calculate

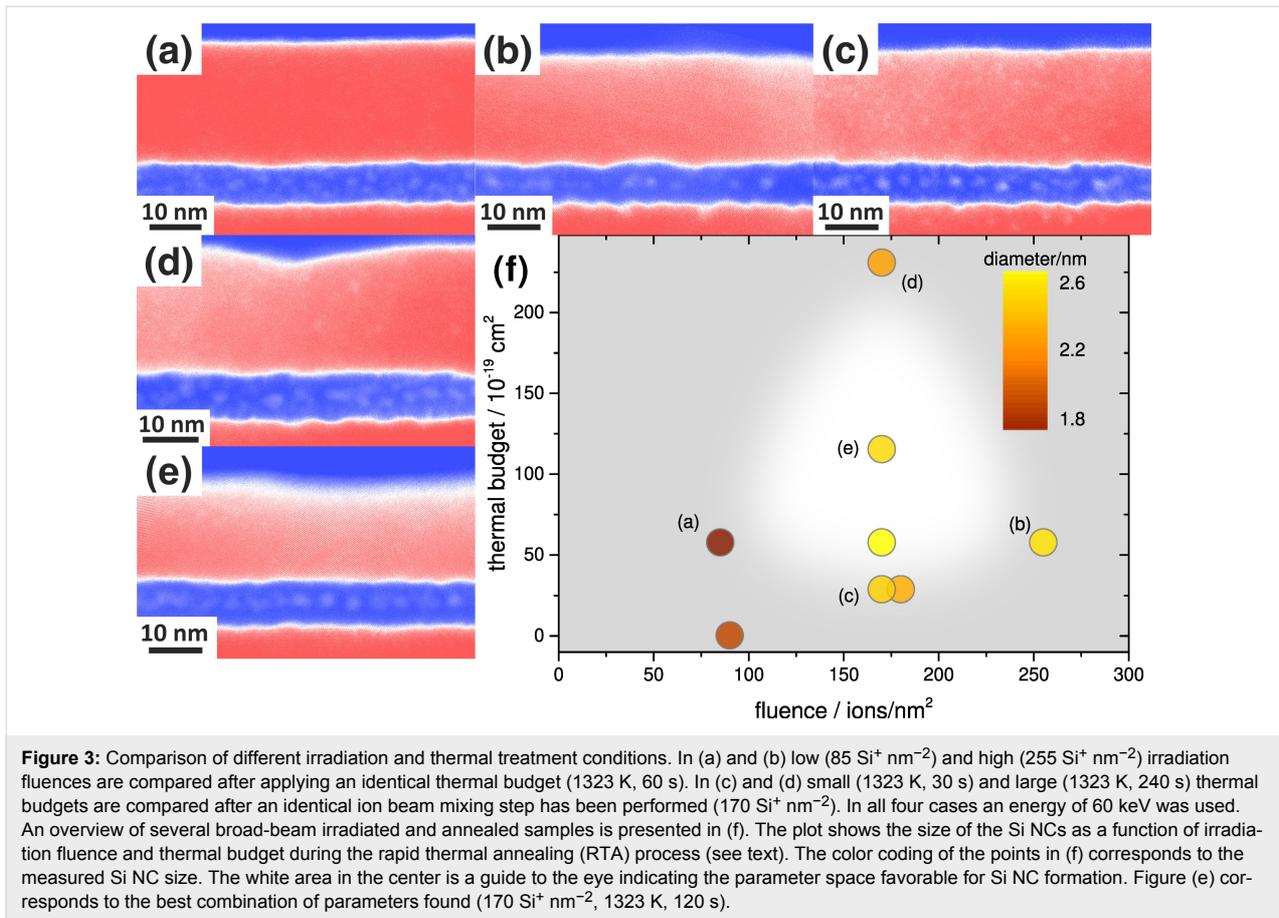

**Figure 3:** Comparison of different irradiation and thermal treatment conditions. In (a) and (b) low (85 Si$^+$ nm$^{-2}$) and high (255 Si$^+$ nm$^{-2}$) irradiation fluences are compared after applying an identical thermal budget (1323 K, 60 s). In (c) and (d) small (1323 K, 30 s) and large (1323 K, 240 s) thermal budgets are compared after an identical ion beam mixing step has been performed (170 Si$^+$ nm$^{-2}$). In all four cases an energy of 60 keV was used. An overview of several broad-beam irradiated and annealed samples is presented in (f). The plot shows the size of the Si NCs as a function of irradiation fluence and thermal budget during the rapid thermal annealing (RTA) process (see text). The color coding of the points in (f) corresponds to the measured Si NC size. The white area in the center is a guide to the eye indicating the parameter space favorable for Si NC formation. Figure (e) corresponds to the best combination of parameters found (170 Si$^+$ nm$^{-2}$, 1323 K, 120 s).

$$\mathrm{TB} = D_0 \int e^{-\frac{E_\mathrm{a}}{k_\mathrm{B} T(t)}} \mathrm{d}t \qquad (1)$$

from the annealing temperature $T$ and the annealing time $t$. The other parameters and constants are the activation energy and the diffusion constant for the dominant diffusion process (SiO diffusion in SiO$_2$) $E_\mathrm{a} = 6.2$ eV and $D_0 = 4 \times 10^4$ cm$^2$/s, respectively, [30] and the Boltzmann constant, $k_\mathrm{B}$.

The effect of different thermal budgets, TB, can be seen from Figure 3c and Figure 3d. Using identical ion beam mixing conditions (170 Si$^+$ nm$^{-2}$ at 60 keV) a too small TB (1323 K, 30 s) results in incomplete (see Figure 3c) or no NC formation while a too high TB (1323 K, 240 s) will result in the decomposition of the earlier formed NCs and an increased interface roughness (see Figure 3d). Complete decomposition requires a higher thermal budget using either longer annealing times or higher temperatures [31] and was observed in this work using simulations.

A wide range of irradiation and annealing conditions has been tested experimentally. An overview of the achieved NC sizes as a function of thermal budget and irradiation fluence is presented in Figure 3f. Some general conclusions can be drawn from this diagram. The smallest NCs can be achieved by using extreme conditions for either fluence (very low, Figure 3a) or thermal budget (very large, Figure 3d). However, from the EFTEM results presented in Figure 3a,d one can see that the NCs formed under these conditions are not well defined and are subject to a rather large size distribution (not shown in this diagram). We conclude that for these cases either insufficient mixing or the onset of NC dissolution, respectively, leads to the observed broad NC size distribution.

The self-assembly of a single δ-layer of unimodal Si NCs in the middle of the 7 nm SiO$_2$ layer requires a fluence of approximately 170 Si$^+$/nm$^2$ and an intermediate thermal budget between $70 \times 10^{-19}$ cm$^2$ and $200 \times 10^{-19}$ cm$^2$. Comparing the best result obtained (Figure 3e) to the predicted mixing profile presented in Figure 2, it is very reasonable that the oxide needs to be Si-enriched to SiO$_{1.5}$. The conditions used in this case were 170 Si$^+$ nm$^{-2}$ with an energy of 60 keV for the mixing and 1323 K, 120 s for the annealing. Under these conditions, a NC diameter of 2.5 nm ± 0.4 nm (the uncertainty is represented by the standard deviation of NC diameter) and an in-plane spacing between the NCs of approximately 12 nm is obtained.

## Ion beam mixing by focused beam irradiation

To realize a single Si NC instead of a 2D array of NCs, the spatial distribution of the mixed material has to be restricted to a volume with a diameter on the order of the NC spacing. To achieve this, the irradiated area has to be downscaled in order to minimize the mixed volume. The minimum mixed volume for a given ion species, energy and target material is related to the size of the collision cascade. Experimentally, this limit can only be achieved if the beam diameter becomes smaller than the lateral straggling in the collision cascade. One particularly flexible way of doing this is to use a focused ion beam. We employed a focused $Ne^+$ beam from a helium ion microscope [32]. Such equipment allows focusing a 30 keV $He^+$ beam to a diameter smaller than 0.5 nm or alternatively a 25 keV $Ne^+$ beam in less than 2 nm. Currently, no other commercially available technology can provide a better-focused ion beam.

However, due to the lateral straggling and the reduced geometrical overlap of neighboring collision cascades, the effective fluence changes with depth. According to Gras-Marti and Sigmund [28], the 1D mixing efficiency for broad-beam irradiation of a layered semi-infinite target is defined as

$$M(x) = \int l^2 \frac{d\sigma(x,l)}{dl} dl \qquad (2)$$

where $x$ denotes the depth along the direction of incidence, $l$ the component of the displacement vector along $x$ of a recoil atom generated at $x$, and $d\sigma/dl$ the differential cross section for the generation of such a recoil. To calculate the 1D mixing efficiency by means of static BCA computer simulation, the number of displacements $N_d$ in a depth interval $\Delta x$ at $x$ with a displacement between $l$ and $l + \Delta l$ is counted for a predefined number $N_i$ of incident ions, being related to the cross section according to

$$N_d(x,l) = n(x)\sigma(x,l)\Delta x N_i \qquad (3)$$

where $n(x)$ denotes the local atomic density. Thus, the mixing efficiency results as

$$M(x) = \frac{1}{n(x)\Delta x N_i} \sum_{\substack{\text{recoils generated} \\ \text{within } \Delta x \text{ around } x}} (x_f - x_i)^2 \qquad (4)$$

where $x_i$ and $x_f$ denote the initial and final depths, respectively, of each recoil generated within $\Delta x$. In Figure 4a the so obtained depth profile of the mixing efficiency of a 25 keV $Ne^+$ broad beam irradiation is plotted.

For line irradiation, with a focused beam along the lateral coordinate $y$, the mixing efficiency varies along $x$ and the lateral coordinate $z$ which is perpendicular to the line direction. Correspondingly, a 2D differential mixing efficiency can be defined as

$$\frac{dM(x,z)}{dz} = \int l^2 \frac{d^2\sigma(x,z,l)}{dz\,dl} dl \qquad (5)$$

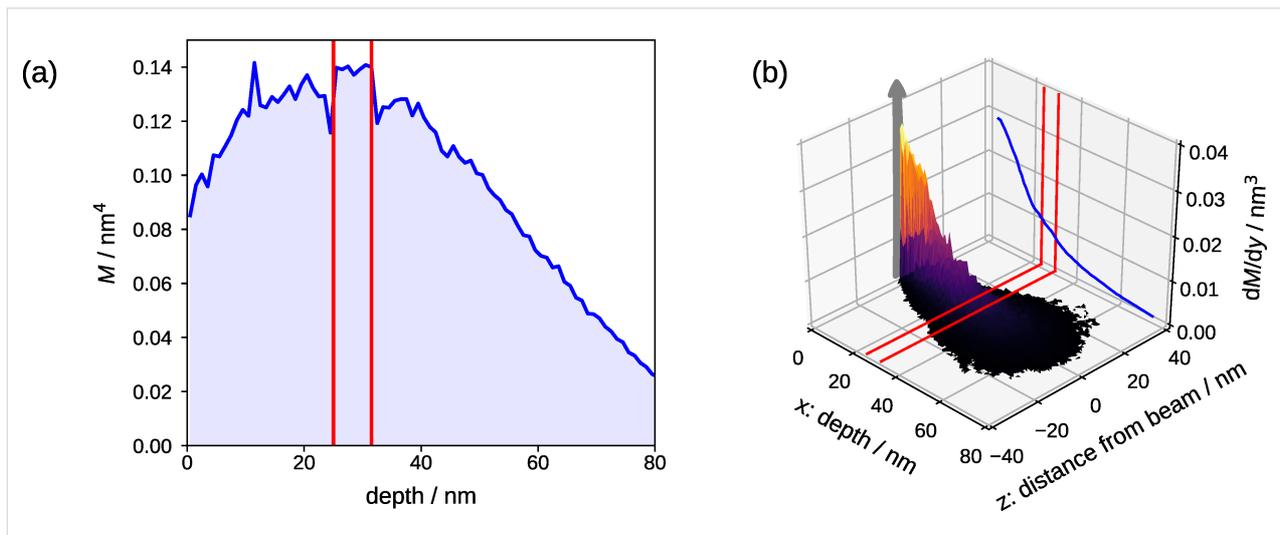

**Figure 4:** Static TRIDYN-based simulation of the mixing efficiency for broad (a) and focused (b) beam irradiation. The graph in (a) shows the mixing efficiency of a 25 keV $Ne^+$ broad beam on a layer stack of 25 nm Si and 6.5 nm $SiO_2$ on a Si substrate. The position of the oxide is indicated by red lines. A similar calculation for a 25 keV focused $Ne^+$ ion beam is presented in (b). The mixing efficiency along the beam direction at the position of the beam is plotted in the $x,y$ plane. The gray arrow indicates the lateral coordinate $y$ along which the beam is scanned. The full width at half maximum of the mixed volume at the depth of the buried oxide is approximately 10 nm. This is five times larger than the beam diameter on the surface of 2 nm.

where d$^2\sigma$/(d$z$d$l$) now denotes the differential cross section of recoil generation at ($x$,$z$) with a displacement $l$ along $x$. Correspondingly, from the BCA simulation,

$$\frac{dM(x,z)}{dz} = \frac{1}{n(x,z)\Delta x \Delta z N_i} \sum_{\substack{\text{recoils generated} \\ \text{within } \Delta x, \Delta z \text{ around } (x,z)}} (x_f - x_i)^2 \quad (6)$$

can be used. The result for this case is given in Figure 4b as the 2D projection of the mixing efficiency for a 25 keV focused Ne$^+$ beam.

While for the broad beam case the mixing efficiency at the surface is slightly smaller than the one at the Si/SiO$_2$ interface and uniform in the lateral plane of the sample, the situation is more complex for the focused beam case. For line irradiation, cascade overlap only happens along the line direction but not perpendicular to it. Due to scattering perpendicular to the line, beam-induced mixing is diluted and the local mixing efficiency decreases with increasing depth and lateral distance from the line position. The lateral distribution of the mixing efficiency on the other hand leads to a concentration of the highly mixed volume in the center of the irradiated line and thus predefines the nucleation site of the NC near the position of the highest Si excess. The maximum mixing efficiency d$M$/d$y$ in the buried oxide – in the center of the line – is approximately 0.01 nm$^3$ (see Figure 4b) which is roughly 1/3 of the mixing efficiency at the surface. For the case of the Ne$^+$ broad beam (Figure 4a), the value is nearly constant over the relevant depth region with a value of ≈0.14 nm$^4$, which is 1.5 times higher than on the surface. It is expected – and confirmed by simulation – that the change in the lateral mixing profile promotes the site-controlled Si NC formation.

To obtain the necessary fully three-dimensional mixing profiles for localized ion beam irradiation, static TRI3DYN [33] simulations have been employed. In Figure 5a, the result for line-profiled focused Ne$^+$ irradiation of 2000 Ne$^+$/nm$^2$ at an energy of 25 keV and a beam diameter of 4 nm is shown. For clarity, only the bottom Si layer as well as the additional Si atoms mixed into the SiO$_2$ layer are displayed.

Taking the results obtained from the BCA-based ion beam mixing simulations as input, kMC simulations can be used to follow the thermally induced phase separation and formation of a single chain of individual Si NCs. Snapshots from the simulation of this process are presented in Figure 5b–d. As one can see, Ostwald ripening leads to a reduction of the NC number with increasing number of Monte Carlo steps, ultimately leading to the formation of a 1D Si NC chain (see Figure 5c). During this process, small clusters decay faster than larger clus-ters which form at the location of the highest Si excess close to the center of the irradiation line. Finally, the NCs start to decay and all the excess Si will be incorporated into the top and bottom Si/SiO$_2$ interfaces.

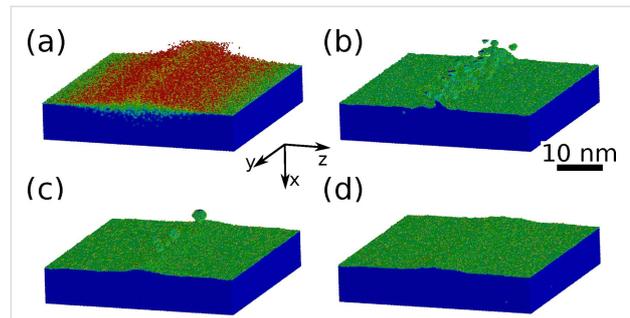

**Figure 5:** Simulation of the formation of a single row of Si NCs by line irradiation. The sample is composed of bulk silicon with a 7 nm thick oxide layer and a 20 nm top silicon layer. The lateral size of the computational cell is 55 nm in both lateral direction ($y$,$z$), with periodic boundary conditions in both lateral directions. (a) Si mixed into SiO$_2$ by a focused Ne$^+$ beam (4 nm beam diameter, 5 nm line width, 2000 Ne$^+$ nm$^{-2}$, 25 keV). The color code corresponds to the binding state of the Si atoms. Blue: Si in the bulk; green: Si at an interface; red: individual Si atoms in an oxide matrix. (b–d) Snapshots from the kMC-based annealing simulation at 1273 K showing the formation ((b) 300 MCs, (c) 3000 MCs), and dissolution ((d) 10000 MCs) of the individual Si NC. The top silicon layer is not shown for clarity.

Two-dimensional (area), 1D (line) and 0D (point-like) irradiations have been carried out using a focused 25 keV Ne$^+$ beam on the aforementioned layer stack. The thicknesses of the layers are 6.5 nm for the buried SiO$_2$ and 25 nm for the top Si. For the 1D irradiation, nominal line widths (widths of the quasi 1D-pattern excluding the beam diameter) ranging from 0 to 6 nm and fluences between 1000 and 5000 Ne$^+$/nm$^2$ have been used. The Ne$^+$ beam diameter was focused to less than 3 nm diameter for all mixing experiments. In combination with the nominal width of the line, this results in an effective irradiated line width of approximately 3–9 nm.

In Figure 6a a cross-sectional Si plasmon-loss filtered TEM image after 1D line irradiation and subsequent RTA is shown. The fluence in the line-pattern is 3000 Ne$^+$/nm$^2$ with a nominal line width of 4 nm. The sample was annealed at 1373 K for 60 s. A Si NC (white), can be seen in the center of the buried SiO$_2$ layer (blue). Note that the NC is aligned with respect to the incident beam. The position of the irradiated line can be deduced from the position of the sputter crater at the surface of the top Si layer (red).

In order to quantify the number of observed Si NCs, a line profile of the Si plasmon-loss intensity is obtained across the layer. This line section is converted into a projected thickness of Si and presented in Figure 6c. Using the intensity of the initial electron beam $I_0$ and the mean free path length (MFPL) $\lambda_{Si}$, we can calculate from the signal intensity $I_{Si}$ the projected silicon thickness

$$d = \ln\left(1 - \frac{I_{Si}}{I_0}\right) \lambda_{Si} . \qquad (7)$$

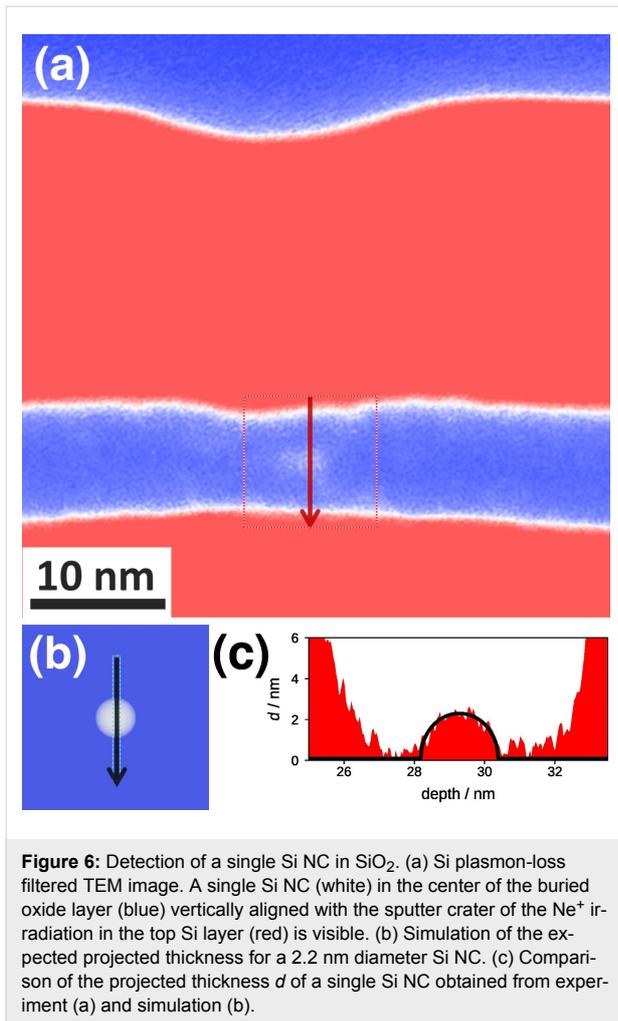

**Figure 6:** Detection of a single Si NC in SiO$_2$. (a) Si plasmon-loss filtered TEM image. A single Si NC (white) in the center of the buried oxide layer (blue) vertically aligned with the sputter crater of the Ne$^+$ irradiation in the top Si layer (red) is visible. (b) Simulation of the expected projected thickness for a 2.2 nm diameter Si NC. (c) Comparison of the projected thickness $d$ of a single Si NC obtained from experiment (a) and simulation (b).

For the inelastically scattered electrons with an energy of interest of 17 ± 2.5 eV, we use $\lambda_{Si}$ = 500 nm and obtain a diameter for the observed Si NC of $d$ = 2.2 nm. The result is presented in Figure 6b and represented by the black line in Figure 6c. The excellent agreement between the measured (Figure 6a) and the expected Si plasmon-loss intensity (Figure 6b) supports the conclusion that only one single Si NC is present in the TEM lamella.

Attempts to further reduce the dimensionality of the irradiation area and perform point irradiations turned out to be impractical due to the necessary increase in Ne$^+$ ion fluence (see Figure 4 and the corresponding discussion). While for the case of line irradiation a fluence of 3000 Ne$^+$/nm$^2$ was sufficient, point irradiation requires a fluence over 10000 Ne$^+$/nm$^2$ in order to achieve a sufficient Si enrichment in the SiO$_2$ layer. Two effects prevent NC self-assembly by point irradiation. First, the sputtering of the top silicon layer becomes significant. Second, Ne bubble formation [34,35] occurred below the SiO$_2$ layer resulting in a strong deformation or even destruction of the SiO$_2$ layer.

## Conclusion

We have demonstrated a novel method for the site-controlled formation of single Si NCs in a buried SiO$_2$ layer using ion beam mixing. This is a promising approach for potential SET devices. Two-dimensional layers and quasi one-dimensional lines of Si NCs have been formed in a buried oxide layer. The optimum thickness of SiO$_2$ for a single δ-layer of NCs was found to be 6.5 ± 1.0 nm. Using ion irradiation of 170 Si$^+$/nm$^2$ at 60 keV and a RTA-induced self-assembly step at 1323 K for 120 s, we achieve Si NCs at the center of the layer with an average spacing of approximately 12 nm and a diameter of 2.5 ± 0.4 nm. Site-controlled formation of single Si NCs of a comparable size was performed by focused Ne$^+$ irradiation with the helium ion microscope. An energy of 25 keV and a fluence of 3000 Ne$^+$/nm$^2$ have been used over a line-shaped irradiation area with a width of 4 nm. After RTA treatment, a single Si NC was isolated and observed in a TEM lamella. This result is corroborated by comparing the simulated and measured Si plasmon-loss signal for a single Si nanocrystal. The NC is centered between the adjacent Si/SiO$_2$ interfaces at the position of the focused beam irradiation and has a diameter of 2.2 nm.

Further minimization of the mixed volume by point-like irradiation with the focused Ne$^+$ ion beam is prohibited by excessive sputtering and Ne bubble formation due to the high fluences required to achieve sufficient mixing. However, the mixed volume can also be reduced by reducing the sample dimensions. This could be achieved for instance by irradiating small pillars whose diameter is on the order of the natural spacing of the Si NCs (slightly larger than 10 nm in this work). This approach would enable broad-beam silicon irradiation and thus avoids Ne bubble formation and increased sputtering in the center of the focused beam. However, such an approach would necessitate a detailed study of the ion-beam-induced erosion of the nanostructured pillars. The required pillar diameter of less than 20 nm is technologically challenging in the context of this fundamental study.

Nevertheless, we demonstrated that ion beam mixing can be used to form layers of Si NCs (broad-beam irradiation) or lines of single Si NCs (focused-beam irradiation) vertically self-centered in a buried oxide layer. The NC size of 2.2 nm and its spacing of 2 nm from the Si/SiO$_2$ interfaces indicates that the so obtained single silicon NC has great potential for the CMOS-compatible integration into future SET applications.

# Experimental
## Computer simulation
Simulation of ion beam mixing was performed using TRIDYN [26] and TRI3DYN [33]. Both programs allow dynamic simulation of the ion beam mixing process taking into account sputtering and accumulation of damage and recoils which lead to a change in stoichiometry. In addition, the latter also allows for complex sample geometries and user defined beam profiles. For the ion beam mixing simulations, only such recoils are assumed to contribute whose start energy exceeds a displacement threshold of 8 eV. A more detailed discussion is given in Möller et al. [36]. The thermally activated phase separation process was simulated with a kinetic Monte Carlo (kMC) code [27].

## Substrate preparation
All substrates used in this study are based on p-doped Si⟨100⟩ wafers with a specific resistivity of 10 Ω cm. The buried SiO$_2$ layer was grown via thermal oxidation at 1123 K in dry O$_2$ atmosphere in a furnace followed by RF-sputtering of an amorphous Si layer. The thickness of the oxide layer was measured by spectroscopic ellipsometry in ambient environment, and the thickness of both layers was confirmed by cross-sectional transmission electron microscopy (TEM).

## Broad-beam irradiation
The irradiation was performed using a 200 kV ion implanter (Danfysik Model 1090) at the Ion Beam Center of the Helmholtz-Zentrum Dresden-Rossendorf (HZDR) [37]. The ion beam incident angle was kept at 7° to avoid channeling. A thermocouple was clipped to the sample during the irradiation to monitor that the temperature does not exceed 423 K. The fluence is measured by a Faraday corner-cup setup.

## Focused beam irradiation
Focused ion beam (FIB) irradiation was performed using a helium ion microscope (HIM) [24,32] (ORION NanoFab, Carl Zeiss). Ne$^+$ ions with an energy of 25 keV were used for ion beam mixing and imaging with either He or Ne was kept to a minimum to avoid additional unintentional damage and mixing. A 10 μm molybdenum aperture was used for Ne$^+$ beam irradiation with spot control number 8. This resulted in a beam current of approximately 150 fA. A double-serpentine scan mode from FIBICS NPVE software was applied to achieve the pattern. The beam diameter was controlled to be better than 3 nm before and after the irradiation using a suitable edge and a 20%/80% criterion.

## Thermal treatment
Rapid thermal processing was conducted with an Allwin21® AccuThermal AW610 tool. All processes were performed in N$_2$ atmosphere with a flow rate of 2.47 L/min (10.0 standard L/min at 2.5 bar gas pressure and room temperature). The ramp-up rate was higher than 25 K/s, thus for simplicity, only the time at peak temperature is taken into account for the thermal budget calculation.

## Energy-filtered transmission electron microscopy (EFTEM)
Cross-sectional samples were obtained by classical lamella preparation including sawing, grinding, polishing, dimpling, and final Ar$^+$ ion milling, as well as FIB (Zeiss NVision 40) milling including a lift-out process. EFTEM images were recorded with an FEI Titan 80-300 microscope using a 5 eV energy-selecting slit at an energy loss of 17 eV, which is the plasmon-loss peak of Si. This energy is preferable, compared to the 25 eV SiO$_2$ plasmon-loss peak, due to its better signal-to-noise ratio. The NC diameter was evaluated using Otsu thresholding [38] in Fiji [39]. The calculation of the projected Si thickness *d* (Equation 7) from the Si plasmon-loss filtered images was performed in the following way. The intensity of a Si plasmon-loss image can be described as

$$I_{\text{loss}}(E, \Delta E) = I_0 \left( 1 - e^{\frac{-d}{\lambda(E, \Delta E)}} \right) \qquad (8)$$

with the characteristic energy-loss *E*, the chosen energy slit width $\Delta E$, the initial intensity $I_0$, and the inelastic mean free path length (MFPL) $\lambda(E,\Delta E)$ for a certain *E* and $\Delta E$. We denote $\lambda(E = 17\text{ eV}, \Delta E = 5\text{ eV})$ as $\lambda_{\text{Si}}$ and, likewise, $I_{\text{loss}}(E = 17\text{ eV}, \Delta E = 5\text{ eV})$ as $I_{\text{Si}}$. We obtain $I_0$ by recording an object-free TEM image in vacuum using the same imaging parameters as for the EFTEM image. In order to determine *d*, we additionally record a zero-loss filtered image in the Si substrate region, which can be expressed by $I_{\text{ZL}} = I_0\exp(-d/\lambda_{\text{in}})$ with $\lambda_{\text{in}}$ the total MFPL for 300 keV electrons in Si. Using $\lambda_{\text{in}} = 180$ nm for 300 keV electrons [25] we can calculate the thickness *d* of the TEM lamella in the Si substrate. By inserting the thickness *d* in Equation 8 one can determine $\lambda_{\text{Si}} = 500$ nm. The knowledge of this value allows us to convert the Si-plasmon-loss filtered image into a projected Si-thickness-map (Equation 7), which is of particular interest for measuring the projected thickness of the Si NCs inside the oxide layer (Figure 6).

# Acknowledgements
XX and TP contributed equally to this manuscript. This work was funded by the European Commission H-2020 programme under grant agreement No. 688072. Daniel Wolf received

*Beilstein J. Nanotechnol.* **2018,** *9,* 2883–2892.

funding from the European Research Council (ERC) under grant No. 715620. Support from Annette Kunz and Romy Aniol for the TEM sample preparation as well as Roman Böttger and the Ion Beam Center at the HZDR for broad-beam irradiation is acknowledged.

## ORCID® iDs

Xiaomo Xu - https://orcid.org/0000-0003-2175-2300
Stefan Facsko - https://orcid.org/0000-0003-3698-3793
Gregor Hlawacek - https://orcid.org/0000-0001-7192-716X